\documentclass[11pt,dvips]{article}
\usepackage{eurosym}
\usepackage{graphicx}
\usepackage{amsmath}
\usepackage{amssymb}
\usepackage{latexsym}
\usepackage{color}
\usepackage{epstopdf}

\input{tcilatex}
\begin{document}

\thispagestyle{empty}

\begin{center}
\vspace{0.7cm}

%%%%%%%%%%%%%%%%%%%%%%%%%%%%%%%%%%%%%%%%%%%%%%%%%%%%%%%%%%%%%%%%%%%%%%%%%%%%%%%%%%%%%%%%%%%%%%%%%%%%%%%%%%%%%%%%%%%%%%%
{\Large \textbf{Unidirectional Gaussian One-Way Steering}}
%%%%%%%%%%%%%%%%%%%%%%%%%%%%%%%%%%%%%%%%%%%%%%%%%%%%%%%%%%%%%%%%%%%%%%%%%%%%%%%%%%%%%%%%%%%%%%%%%%%%%%%%%%%%%%%%%%%%%

\vspace{0.5cm}\textbf{Jamal El Qars}

\textit{Department of Physics, Faculty of Applied Sciences, Ait-Melloul, Ibn
Zohr University, Agadir, Morocco}

\vspace{0.9cm}

\vspace{0.9cm}\textbf{Abstract}
\end{center}

Steering is a type of quantum nonlocality that exhibits an inherent asymmetry
between two observers. In a nondegenerate three-level laser coupled to a
two-mode squeezed vacuum reservoir, we examine, under realistic experimental
conditions, the Gaussian steering of two laser modes, $\mathcal{A}$ and $\mathcal{%
B}$, generated within the cascade transitions, respectively. We find that
the $\mathcal{A}\rightarrow \mathcal{B}$ steerability is always higher than
that from $\mathcal{B}\rightarrow \mathcal{A}$; in addition, the steering asymmetry
cannot exceed $\ln 2$, which implies that the state $\hat{\varrho}_{\mathcal{AB}}$
never diverges to an extremal asymmetry state. We show how squeezed noise
can play a constructive role in realizing one-way steering. As the main result,
we demonstrate that the state $\hat{\varrho}_{\mathcal{AB}}$ can exhibit one-way
steering solely from $\mathcal{A}\rightarrow \mathcal{B}$, which we show to
emerge as a consequence of the fact that the intensity difference of the
modes $\mathcal{A}$ and $\mathcal{B}$ is verified to remain always positive, irrespective of the physical and environmental parameters of $\hat{\varrho}_{%
\mathcal{AB}}$. The generated unidirectional one-way steering may provide a
useful resource for the distribution of the trust in future asymmetric
quantum information tasks.

\section{Introduction}

In their famous 1935 paper \cite{EPR}, Einstein, Podolsky, and Rosen (EPR) highlighted that when two particles are in a pure entangled state, a
suitable measurement performed on one particle induces an apparent nonlocal
collapse of the wave function of the other. Based on this argument, which is
often referred to as the \textquotedblleft EPR paradox,\textquotedblright the authors argued that the notion of local causality could not be compatible with the completeness of quantum mechanics \cite{EPR}. In his
response to the EPR paper, Schr\"{o}dinger \cite{Erwin} introduced the
concept of \textit{steering} as a nonlocal quantum effect by which sets of
quantum states can be remotely prepared via local measurements.

In terms of the violations of local hidden state \cite{Wiseman}, steering has
been conscientiously defined as an intermediate form of non-separable quantum
correlations, which is stronger than entanglement \cite{QE}, but weaker than
Bell-nonlocality \cite{Bell}. From a quantum information perspective,
EPR-steering describes the ability that allows an observer (say Alice) to
apparently adjust (i.e., \textit{to steer}) the state of another distant observer (Bob) via local measurements, by exploiting their shared entanglement
\cite{Wiseman}. However, such correlations allow the verification of
entanglement distribution, even if the measurement devices of one observer
are untrusted \cite{Jones,Bown}. In other words, if Alice and Bob share a
bipartite state $\hat{\varrho}_{\mathrm{AB}}$, which is at least steerable in
one direction (e.g., from $\mathrm{A\rightarrow B}$), Alice
can convince Bob, who does not trust Alice, that their shared state $\hat{%
\varrho}_{\mathrm{AB}}$ is entangled by performing local measurements and
classical communications \cite{Wiseman}. Owing to its intriguing feature, EPR
steering has recently garnered considerable attention in quantum optics
and quantum information communities \cite{OWS}.

Based on the inferred Heisenberg uncertainty principal, Reid proposed an
experimental criterion for capturing the essence of the EPR-paradox in a
continuous-variable setting \cite{Reid}, which was later proven to be a
necessary and sufficient condition for detecting Gaussian steerable states
under Gaussian measurements \cite{Wiseman}. The first experimental
demonstration of steering was achieved in \cite{Ou}, and was followed by a significant number of experiments \cite{Exp steering}. To quantify the amount by
which a bipartite Gaussian state $\hat{\varrho}_{\mathcal{XY}}$ is steerable
under Gaussian measurements, Kogias et al. \cite{Kogias} derived a
computable measure defined by the means of the R\'{e}nyi-2 entropy, and they determined that for generic two-mode Gaussian states (TMGSs), the proposed
measure reduces to a form of coherent information, which was proven never to
exceed entanglement, but confined to it on pure states \cite{Kogias}.

A characteristic trait of steering that distinguishes it from entanglement and
Bell-nonlocality is the asymmetry between the observers, Alice and Bob, i.e.,
their shared state $\hat{\varrho}_{\mathrm{AB}}$ may be steerable in one
direction (say, e.g., from $\mathrm{A}\rightarrow \mathrm{B}$) but not vice
versa, which is referred to as one-way steering \cite{onews}. Because the
efficiency of asymmetric quantum information tasks, such as one-sided
device-independent quantum key distribution \cite{QKD}, subchannel
discrimination \cite{SD}, universal one-way quantum computing \cite{CMLi},
secure quantum teleportation \cite{sqt}, and quantum secret sharing \cite%
{Kogias2}, significantly depends on the direction of the measurement, it is widely believed that the key ingredient in such protocols is the
one-way steering phenomenon \cite{onews}.

In the context of quantum optics \cite{S-Z}, nondegenerate three-level
lasers are demonstrated to exhibit various nonclassical features, such as quenching
of spontaneous emissions and quadrature-squeezing \cite{Scully',Blockly},
which can be used to investigate different aspects of quantum nonlocality
\cite{Han,Qamar,TesfaOpt}. A nondegenerate three-level laser is a quantum
optical system in which a set of three-level atoms, in a cascade
configuration, are injected at a constant rate inside a resonant cavity \cite%
{S-Z}. When a single atom transits from the upper to lower level via the
intermediate level, two strongly correlated photons are generated \cite%
{Rempe}. If the two emitted photons are of different frequencies, the laser
is regarded a nondegenerate three-level laser \cite{NDCEL}; otherwise, it is called a
degenerate three-level laser \cite{DCEL}. In such lasers, the
fundamental role is facilitated by atomic coherence, which can be induced by first preparing the atoms in a coherent superposition of the upper and
lower levels (injected coherence) \cite{Blockly} or coupling the same levels using a strong classical field (driven coherence) \cite{Han,NDCEL}, or adopting the two mechanisms simultaneously \cite{Ansari4686}.

Although entanglement has been comprehensively studied in nondegenetae three-level
cascade lasers \cite{Han,NDCEL,QENDL}, only a few studies have been conducted in the past years to examine Gaussian steering in such systems. For
instance, based on the measure of Kogias et al. \cite{Kogias}, Ullah et al. \cite{Ullah} studied dynamical Gaussian steering in a
cascade laser coupled to a vacuum reservoir, where the atomic coherence were introduced via the driven coherence process. However, Zhong et al. \cite{LPL} considered a similar system with an injected coherence process, and under a thermal effect, they investigated stationary Gaussian steering by employing
the Reid criterion \cite{Reid}. To the best of our knowledge, no previous study considering a nondegenerate three-level laser has analyzed Gaussian
steering under squeezing effect.

Here, we theoretically analyze quantum steering in a TMGS $\hat{\varrho}_{%
\mathcal{AB}}$ coupled to a two-mode squeezed vacuum, where the modes $%
\mathcal{A}$ and $\mathcal{B}$ are, respectively, generated within the
cascade transitions of a nondegenerate three-level laser. Using realistic
experimental parameters from \cite{Han,Meschede}, we demonstrate that stationary
asymmetric steering can be generated between the laser modes $\mathcal{A}$
and $\mathcal{B}$. In addition, by altering various practically accessible
parameters, we show that one-way steering in the state $\hat{\varrho}_{%
\mathcal{AB}}$ can be achieved.

Unlike various TMGS $\hat{\varrho}_{\mathcal{XY}}$ systems \cite%
{LPL,SOPTO,magno,LBS,ElQars,PRA,kongHu} in which one-way steering is bidirectional, i.e., it can emerge from $\mathcal{X}\rightarrow\mathcal{%
Y}$ and $\mathcal{Y}\rightarrow\mathcal{X}$, herein, we mainly demonstrate that the state at hand $\hat{\varrho}_{\mathcal{AB}}$ can solely exhibit
one-way steering from $\mathcal{A}\rightarrow\mathcal{B}$. Furthermore, we
show that this unidirectionality restriction emerges as a consequence of the
fact that the intensity difference of the modes $\mathcal{A}$ and $\mathcal{B%
}$ is verified to always remain positive, regardless of the physical and
environmental parameters of $\hat{\varrho}_{\mathcal{AB}}$. From a practical perspective, the unidirectional $\mathcal{A}\rightarrow \mathcal{B}$
one-way steering traduces the existence of situations in which Alice, alone,
can influence Bob's state via her choice of measurement basis, while a similar operation in the reverse direction is, in general, impossible, which
may facilitate novel perspectives in one-way quantum communication and computation
protocols \cite{sqt,QSS,RQC}.

Our work is partly motivated by the considerable attention recently garnered by the one-way steering phenomenon as the key ingredient in the implementation of asymmetric quantum information protocols \cite{OWS}.
And partly by the centrality of Gaussian states in quantum information
processing owing to their peculiar structural properties that make their
theoretical description amenable to an analytical analysis, including their ability to generate, control, and measure such states with notable efficiency
in feasible experimental settings \cite{Shapiro}.

The remainder of this paper is organized as follows. In Sec. \ref{SecII}, we
introduce our model and derive the master equation of the considered state $\hat{\varrho}_{\mathcal{AB}}$. Next, we obtain an analytical formula of the covariance matrix describing the Gaussian stationary state of the
laser modes $\mathcal{A}$ and $\mathcal{B}$. In Sec. \ref{SecIII}, using the
measure proposed by Kogias et al. \cite{Kogias}, we quantify the $\mathcal{A}%
\rightarrow \mathcal{B}$ and $\mathcal{B}\rightarrow \mathcal{A}$
steerabilities and study their behaviors under practical experimental
conditions. Furthermore, we calculate the explicit expression of the intensity
difference of the modes $\mathcal{A}$ and $\mathcal{B}$ and verify it to
remain always positive, regardless of the physical and environmental
parameters of the state $\hat{\varrho}_{\mathcal{AB}}$. Finally, in Sec. \ref%
{SecIV}, we draw our conclusions.

\section{Model and master equation}

\label{SecII}

In a doubly-resonant cavity coupled to a two-mode squeezed vacuum reservoir,
we consider an ensemble of nondegenerate three-level atoms in a cascade
configuration \cite{S-Z,Banacloche}. The atoms are assumed to be injected
into the cavity at a rate $r_{0}$, and then removed after a certain time $\tau $
during which a single atom interacts \textit{resonantly} with two bosonic
modes of the quantized cavity field \cite{Fesseha}. The $j\text{th}$ cavity
mode is characterized by its annihilation operator $\hat{a}_{j}$, frequency $%
\omega _{j}$, and cavity decay rate $\kappa _{j}$. We adopt $%
|a\rangle$, $|b\rangle$, and $|c\rangle $ to denote the top, intermediate, and bottom
levels of a single three-level atom, respectively. Furthermore, we assume that the atoms
are initially prepared in an arbitrary coherent superposition of the top $%
|a\rangle $ and bottom $|c\rangle $ levels, with the populations $|\wp
_{a}|^{2}$ and $|\wp _{c}|^{2}$ \cite{CS}. Accordingly, the initial
state of a single atom and its corresponding density operator are
respectively \cite{S-Z}
\begin{eqnarray}
|\psi _{\mathrm{a}}^{(0)}\rangle &=&\wp _{a}|a\rangle +\wp _{c}|c\rangle ,
\label{as} \\
\hat{\varrho}_{\mathrm{a}}^{(0)} &=&\varrho _{aa}^{(0)}|a\rangle \langle
a|+\varrho _{ac}^{(0)}|a\rangle \langle c|+\varrho _{ac}^{(0)\ast }|c\rangle
\langle a|+\varrho _{cc}^{(0)}|c\rangle \langle c|,  \label{rdo}
\end{eqnarray}%
where $\varrho _{aa}^{(0)}=|\wp _{a}|^{2}$ and $\varrho _{cc}^{(0)}=|\wp
_{c}|^{2}$ represent the top and bottom level initial populations, respectively, and $\varrho
_{ac}^{(0)}=\varrho _{ca}^{(0)\ast }=\wp _{a}\wp _{c}^{\ast }$ is the
initial two-photon atomic coherence.
\begin{figure}[tbh]
\centerline{\includegraphics[width=10.5cm]{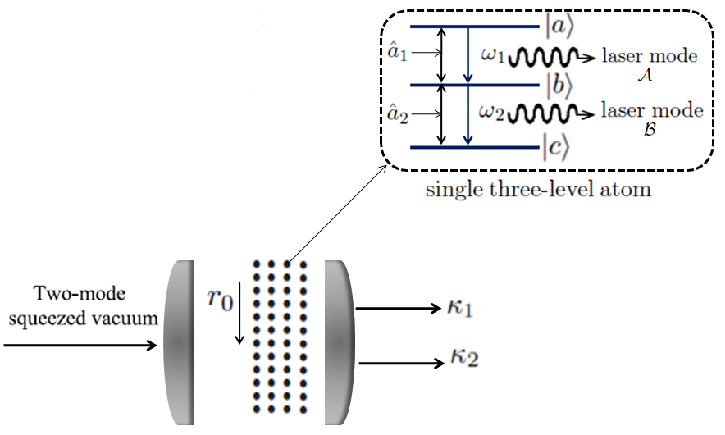}}
\caption{Schematic diagram of a nondegenerate three-level laser coupled to a
two-mode squeezed vacuum reservoir. $r_{0}$ denotes the rate at which a set of
nondegenerate three-level atoms in a cascade configuration are injected into
the cavity. $|a\rangle$, $|b\rangle$, and $|c\rangle$ represent the
top, intermediate, and bottom energy levels of a single three-level atom, respectively.
The dipole-allowed transitions $|a\rangle \rightarrow |b\rangle $ and $%
|b\rangle \rightarrow |c\rangle$ are assumed to be resonant with two cavity
modes $\hat{a}_{1}$ and $\hat{a}_{2}$; however, the transition $|a\rangle
\rightarrow |c\rangle $ is dipole-forbidden. $\protect\omega_{1}$ and $%
\protect\kappa_{1}$($\protect\omega _{2}$ and $\protect\kappa_{2}$) respectively denote the frequency and the cavity decay rate of the laser mode $%
\mathcal{A}$($\mathcal{B}$) generated during the transition $|a\rangle
\rightarrow |b\rangle $($|b\rangle \rightarrow |c\rangle $).}
\label{f0}
\end{figure}

In the interaction picture, under the rotating-wave approximation, the
interaction of a single three-level atom with the two cavity modes can be
expressed by the Hamiltonian \cite{S-Z}
\begin{equation}
\mathcal{\hat{H}}_{\mathrm{int}}=\mathrm{i}\hbar g(\hat{a}_{1}|a\rangle
\langle b|+\hat{a}_{2}|b\rangle \langle c|-|b\rangle \langle a|\hat{a}%
_{1}^{\dag }-|c\rangle \langle b|\hat{a}_{2}^{\dag }),  \label{HM}
\end{equation}%
where the coupling constant $g$ is assumed to be the same for both
transitions $|a\rangle \rightarrow |b\rangle $ and $|b\rangle \rightarrow
|c\rangle $ \cite{S-Z,Fesseha}.

Subsequently, based on the procedure developed in \cite{Louisell,Sargent}, we derive
the master equation of the reduced density operator $\hat{\varrho}_{\mathcal{%
AB}}$ for the two-mode laser $\mathcal{A}$ and $\mathcal{B}$. First, using $\hat{\varrho}_{\left( \mathrm{a,}\mathcal{AB}\right)
}(t,t_{j}) $, we represent the density operator at time $t$ for the two laser modes plus a
single atom injected in the cavity at an earlier time $t_{j}$. The density
operator describing an ensemble of atoms in the cavity plus the modes $%
\mathcal{A}$ and $\mathcal{B}$ at time $t$ can be expressed as
\begin{equation}
\hat{\varrho}_{(\mathrm{a},\mathcal{AB})}(t)=r_{0}\dsum\limits_{j}\hat{%
\varrho}_{(\mathrm{a},\mathcal{AB})}(t,t_{j})\Delta t,  \label{DS}
\end{equation}%
where $r_{0}\Delta t$ denotes the total number of atoms pumped into the cavity
within a short time interval $\Delta t$.

In the limit $\Delta t\rightarrow 0$, the summation can be replaced by
integration, i.e.,
\begin{equation}
\hat{\varrho}_{(\mathrm{a},\mathcal{AB})}(t)=r_{0}\int_{t-\tau }^{t}\hat{%
\varrho}_{(\mathrm{a},\mathcal{AB})}(t,t^{\prime })dt^{\prime },  \label{DME}
\end{equation}
where $\tau $ denotes the time after which the atoms leave the cavity with $%
t-\tau \leqslant t^{\prime }\leqslant t.$ Next, by differentiating both
sides of Eq. (\ref{DME}) relative to $t$, we have
\begin{equation}
\frac{d\hat{\varrho}_{(\mathrm{a},\mathcal{AB})}(t)}{dt}=r_{0}\frac{d}{dt}%
\int_{t-\tau }^{t}\hat{\varrho}_{(\mathrm{a},\mathcal{AB})}(t,t^{\prime
})dt^{\prime },  \label{MS1}
\end{equation}%
which can be rewritten using the Leibnitz rule as
\begin{equation}
\frac{d\hat{\varrho}_{(\mathrm{a},\mathcal{AB})}(t)}{dt}=r_{0}\left[ \hat{%
\varrho}_{(\mathrm{a},\mathcal{AB})}(t,t)-\hat{\varrho}_{(\mathrm{a},%
\mathcal{AB})}(t,t-\tau )\right] +r_{0}\int_{t-\tau }^{t}\frac{\partial \hat{%
\varrho}_{(\mathrm{a},\mathcal{AB})}(t,t^{\prime })}{\partial t}dt^{\prime }.
\label{Rule}
\end{equation}

Further, we assume that the atomic states and laser modes are uncorrelated
at the instant $t$ in which the atom is pumped into the cavity (i.e., the
Markov approximation), then $\hat{\varrho}_{(\mathrm{a},\mathcal{AB})}(t,t)=%
\hat{\varrho}_{\mathcal{AB}}(t)\hat{\varrho}_{\mathrm{a}}(0)$ \cite{Gardiner}%
. Moreover, with the assumption that the atomic states and laser modes are
uncorrelated just after the atom left the cavity, i.e., $\hat{\varrho}_{(%
\mathrm{a},\mathcal{AB})}(t,t-\tau )=\hat{\varrho}_{\mathcal{AB}}(t)\hat{%
\varrho}_{\mathrm{a}}(t-\tau )$, Eq. (\ref{Rule}) becomes
\begin{equation}
\frac{d\hat{\varrho}_{(\mathrm{a},\mathcal{AB})}(t)}{dt}=r_{0}\left[ \hat{%
\varrho}_{\mathrm{a}}(0)-\hat{\varrho}_{\mathrm{a}}(t-\tau )\right] \hat{%
\varrho}_{\mathcal{AB}}(t)+r_{0}\int_{t-\tau }^{t}\frac{\partial \hat{\varrho%
}_{(\mathrm{a},\mathcal{AB})}(t,t^{\prime })}{\partial t}dt^{\prime },
\label{MS2}
\end{equation}%
where $\hat{\varrho}_{\mathrm{a}}(t-\tau )$ represents the density operator for an
atom injected at $t-\tau $ and $\hat{\varrho}_{\mathrm{a}}(0)$ is the
initial density operator for each atom given by Eq. (\ref{rdo}). Hereinafter, we use $\hat{\varrho}_{\mathcal{AB}}(t)\equiv \hat{\varrho}(t)$ for
simplicity of notation.

However, using Eq. (\ref{DME}) and the fact that $\frac{\partial }{%
\partial t}\hat{\varrho}_{(\mathrm{a},\mathcal{AB})}(t,t^{\prime })=\frac{-%
\mathrm{i}}{\hbar }[\mathcal{\hat{H}}_{\mathrm{int}},\hat{\varrho}_{(\mathrm{%
a},\mathcal{AB})}(t,t^{\prime })]$, Eq. (\ref{MS2}) is expressed as:
\begin{equation}
\frac{d\hat{\varrho}_{(\mathrm{a},\mathcal{AB})}(t)}{dt}=r_{0}\left[ \hat{%
\varrho}_{\mathrm{a}}(0)-\hat{\varrho}_{\mathrm{a}}(t-\tau )\right] \hat{%
\varrho}(t)-\frac{\mathrm{i}}{\hbar }[\mathcal{\hat{H}}_{\mathrm{int}},\hat{%
\varrho}_{(\mathrm{a},\mathcal{AB})}(t)].  \label{MS3}
\end{equation}

Because we are solely interested in the dynamics of $\hat{\varrho}(t)$, we
trace both sides of Eq. (\ref{MS3}) over the atomic variables to obtain
\begin{align}
\frac{d\hat{\varrho}(t)}{dt}& =\frac{-i}{\hbar }\mathrm{Tr}_{\mathrm{a}}[%
\mathcal{\hat{H}}_{\mathrm{int}},\hat{\varrho}_{(\mathrm{a},\mathcal{AB}%
)}]+\sum\limits_{j=1,2}\left( \frac{\kappa_{j} (N+1)}{2}\mathcal{L}\left[
\hat{a}_{j}\right] \hat{\varrho}+\frac{\kappa_{j} N}{2}\mathcal{L}[\hat{a}%
_{j}^{\dag }]\hat{\varrho}\right)  \notag \\
& -\sum\limits_{j\neq k=1,2}\left( \kappa_{j} M^{\ast }\mathcal{D}\left[
\hat{a}_{j},\hat{a}_{k}\right] \hat{\varrho}+\kappa_{j} M\mathcal{D}[\hat{a}%
_{j}^{\dag },\hat{a}_{k}^{\dag }]\hat{\varrho}\right) ,  \label{MS4}
\end{align}
where we used $\mathrm{Tr}_{\mathrm{a}}\hat{\varrho}_{\mathrm{a}}(0)=\mathrm{%
Tr}_{\mathrm{a}}\hat{\varrho}_{\mathrm{a}}(t-\tau )=1$ and $\mathrm{Tr}_{%
\mathrm{a}}\hat{\varrho}_{(\mathrm{a},\mathcal{AB})}(t)=\hat{\varrho}(t)$.

In Eq. (\ref{MS4}), the two last terms are added to account for the damping
of the laser modes through the squeezed vacuum reservoir, where $\mathcal{L}%
\left[ \hat{a}_{j}\right] \hat{\varrho}=2\hat{a}_{j}\hat{\varrho}\hat{a}%
_{j}^{\dag }-\hat{a}_{j}^{\dag }\hat{a}_{j}\hat{\varrho}-\hat{\varrho}\hat{a}%
_{j}^{\dag }\hat{a}_{j}$ and $\mathcal{D}[\hat{a}_{j},\hat{a}_{k}]\hat{%
\varrho}=2\hat{a}_{j}\hat{\varrho}\hat{a}_{k}-\hat{a}_{k}\hat{a}_{j}\hat{%
\varrho}-\hat{\varrho}\hat{a}_{k}\hat{a}_{j}$, $N=\sinh ^{2}\left( r\right) $
and $M=\sinh \left( r\right) \cosh \left( r\right) \exp (i\phi )$ with $r$
and $\phi $ denoting the squeezing parameter and phase of the
squeezed vacuum, respectively \cite{Tanas}. We note that Gilles et al. \cite{Gilles} proposed an experimentally realizable scheme for coupling two cavity
modes to a two-mode squeezed vacuum using a lossy port mirror. Furthermore,
they demonstrated how the intra-cavity squeezing can be effectively enhanced.

Now, substituting Eq. (\ref{HM}) in Eq. (\ref{MS4}), and performing the trace
operation, we deduce that
\begin{eqnarray}
\frac{d\hat{\varrho}(t)}{dt} &=&g(\hat{\varrho}_{ab}\hat{a}_{1}^{\dag }-\hat{%
a}_{1}^{\dag }\hat{\varrho}_{ab}+\hat{\varrho}_{bc}\hat{a}_{2}^{\dag }-\hat{a%
}_{2}^{\dag }\hat{\varrho}_{bc}+\hat{a}_{1}\hat{\varrho}_{ba}-\hat{\varrho}%
_{ba}\hat{a}_{1}+\hat{a}_{2}\hat{\varrho}_{cb}-\hat{\varrho}_{cb}\hat{a}%
_{2})+  \notag \\
&&\sum\limits_{j=1,2}\left( \frac{\kappa _{j}(N+1)}{2}\mathcal{L}\left[ \hat{%
a}_{j}\right] \hat{\varrho}+\frac{\kappa _{j}N}{2}\mathcal{L}[\hat{a}%
_{j}^{\dag }]\hat{\varrho}\right) -\sum\limits_{j\neq k=1,2}\left( \kappa
_{j}M^{\ast }\mathcal{D}\left[ \hat{a}_{j},\hat{a}_{k}\right] \hat{\varrho}%
+\kappa _{j}M\mathcal{D}[\hat{a}_{j}^{\dag },\hat{a}_{k}^{\dag }]\hat{\varrho%
}\right) ,  \label{MS5}
\end{eqnarray}%
where $\hat{\varrho}_{mn}=\langle m|\hat{\varrho}_{(\mathrm{a},\mathcal{AB}%
)}|n\rangle $ for $m$, $n=a,b,c$.

The next step is to obtain the density operators $\hat{\varrho}_{ab}=\langle
a| \hat{\varrho}_{(\mathrm{a},\mathcal{AB})}| b\rangle $ and $\hat{\varrho}%
_{bc}=\langle b|\hat{\varrho}_{(\mathrm{a},\mathcal{AB})}| c\rangle $ that
appear in Eq. (\ref{MS5}). Accordingly, we multiply Eq. (\ref{MS3}) on the
right by $|n\rangle $ and on the left by $\langle m|$, and further assume
that the atoms decay to energy levels other than the three levels $|a\rangle
$, $|b\rangle $, or $|c\rangle$ when they leave the cavity, i.e., $\langle
m\left\vert \hat{\varrho}_{\mathrm{a}}(t-\tau )\right\vert n\rangle =0$ for $m,n=a,b,c$. Hence, we obtain
\begin{equation}
\frac{d\hat{\varrho}_{mn}}{dt}=r_{0}\langle m\left\vert \hat{\varrho}_{%
\mathrm{a}}(0)\right\vert n\rangle \hat{\varrho}(t)-\frac{\mathrm{i}}{\hbar }%
\langle m|[\mathcal{\hat{H}}_{\mathrm{int}},\hat{\varrho}_{(\mathrm{a},%
\mathcal{AB})}]|n\rangle -\gamma _{mn}\hat{\varrho}_{mn},  \label{MS6}
\end{equation}
where the term $\gamma _{mn}$ is introduced to account for the spontaneous
emission and dephasing processes \cite{S-Z}.

Therefore, using Eq. (\ref{MS6}) together with Eqs. (\ref{rdo}) and (\ref{HM}%
), we obtain
\begin{eqnarray}
\frac{d\hat{\varrho}_{ab}}{dt} &=&g(\hat{\varrho}_{ac}\hat{a}_{2}^{\dag }+%
\hat{a}_{1}\hat{\varrho}_{bb}-\hat{\varrho}_{aa}\hat{a}_{1})-\gamma _{ab}%
\hat{\varrho}_{ab},  \label{MS7} \\
\frac{d\hat{\varrho}_{bc}}{dt} &=&g(\hat{a}_{2}\hat{\varrho}_{cc}-\hat{%
\varrho}_{bb}\hat{a}_{2}-\hat{a}_{1}^{\dag }\hat{\varrho}_{ac})-\gamma _{bc}%
\hat{\varrho}_{bc},  \label{MS8} \\
\frac{d\hat{\varrho}_{ac}}{dt} &=&r_{0}\varrho _{ac}^{(0)}\hat{\varrho}+g(%
\hat{a}_{1}\hat{\varrho}_{bc}-\hat{\varrho}_{ab}\hat{a}_{2})-\gamma _{ac}%
\hat{\varrho}_{ac},  \label{MS9} \\
\frac{d\hat{\varrho}_{aa}}{dt} &=&r_{0}\varrho _{aa}^{(0)}\hat{\varrho}+g(%
\hat{\varrho}_{ab}\hat{a}_{1}^{\dag }+\hat{a}_{1}\hat{\varrho}_{ba})-\gamma
_{a}\hat{\varrho}_{aa},  \label{MS10} \\
\frac{d\hat{\varrho}_{bb}}{dt} &=&g(\hat{\varrho}_{bc}\hat{a}_{2}^{\dag }+%
\hat{a}_{2}\hat{\varrho}_{cb}-\hat{a}_{1}^{\dag }\hat{\varrho}_{ab}-\hat{%
\varrho}_{ba}\hat{a}_{1})-\gamma _{b}\hat{\varrho}_{bb},  \label{MS11} \\
\frac{d\hat{\varrho}_{cc}}{dt} &=&r_{0}\varrho _{cc}^{(0)}\hat{\varrho}-g(%
\hat{a}_{2}^{\dag }\hat{\varrho}_{bc}+\hat{\varrho}_{cb}\hat{a}_{2})-\gamma
_{c}\hat{\varrho}_{cc},  \label{MS12}
\end{eqnarray}%
where $\gamma _{j}$ ($j=a,b,c$) represents the $j$\textrm{th} atomic-level
spontaneous emission decay rate, $\gamma _{ac}$ denotes the two-photon dephasing
rate, while $\gamma _{ab}$ and $\gamma _{bc}$ are the single-photon
dephasing rates \cite{Anom}.

In the good cavity limit, i.e., $\kappa _{j}\ll \gamma _{j}$, the adiabatic approximation scheme can be performed \cite{S-Z}, where the time
derivative in Eqs. [(\ref{MS9})--(\ref{MS12})] can be set to zero. Moreover,
the linear approximation \cite{S-Z,Louisell,Sargent}, allows us to drop the
terms containing $g$ in the same equations. Accordingly, we obtain
\begin{eqnarray}
\hat{\varrho}_{aa} &=&\frac{r_{0}\varrho _{aa}^{(0)}}{\gamma }\hat{\varrho},%
\text{ }  \label{aa} \\
\text{\ }\hat{\varrho}_{bb} &=&0,  \label{bb} \\
\text{ }\hat{\varrho}_{cc} &=&\frac{r_{0}\varrho _{cc}^{(0)}}{\gamma }\hat{%
\varrho},\text{ }  \label{cc} \\
\text{\ }\hat{\varrho}_{ac} &=&\frac{r_{0}\varrho _{ac}^{(0)}}{\gamma }\hat{%
\varrho},  \label{ac}
\end{eqnarray}%
where we used $\gamma _{a}=\gamma _{b}=\gamma _{c}=\gamma
_{ab}=\gamma _{ac}=\gamma _{bc}\equiv \gamma$, for simplicity.

Next, by combining Eqs. [(\ref{aa})--(\ref{ac})] with Eqs. (\ref{MS7}) and (%
\ref{MS8}) and applying the adiabatic approximation again, we obtain
\begin{eqnarray}
\hat{\varrho}_{ab} &=&\frac{gr_{0}}{\gamma ^{2}}\left( \varrho _{ac}^{(0)}%
\hat{\varrho}\hat{a}_{2}^{\dag }-\varrho _{aa}^{(0)}\hat{\varrho}\hat{a}%
_{1}\right) ,\text{ }  \label{ab} \\
\text{\ }\hat{\varrho}_{bc} &=&\frac{gr_{0}}{\gamma ^{2}}\left( \varrho
_{cc}^{(0)}\hat{a}_{2}\hat{\varrho}-\varrho _{ac}^{(0)}\hat{a}_{1}^{\dag }%
\hat{\varrho}\right) .  \label{bc}
\end{eqnarray}

Finally, using balanced cavity losses, i.e., $\kappa _{1}=\kappa _{2}\equiv
\kappa $, and substituting Eqs. (\ref{ab}) and (\ref{bc}) into Eq. (\ref{MS5}%
), we obtain the master equation of the reduced density operator $\hat{%
\varrho}(t)$ for the laser modes $\mathcal{A}$ and $\mathcal{B}$
\begin{eqnarray}
\frac{d\hat{\varrho}(t)}{dt} &=&\frac{\kappa (N+1)}{2}[2\hat{a}_{1}\hat{%
\varrho}\hat{a}_{1}^{\dag }-\hat{a}_{1}^{\dag }\hat{a}_{1}\hat{\varrho}-\hat{%
\varrho}\hat{a}_{1}^{\dag }\hat{a}_{1}]+\frac{\kappa N}{2}[2\hat{a}%
_{2}^{\dag }\hat{\varrho}\hat{a}_{2}-\hat{a}_{2}\hat{a}_{2}^{\dag }\hat{%
\varrho}-\hat{\varrho}\hat{a}_{2}\hat{a}_{2}^{\dag }]+  \notag \\
&&\frac{1}{2}\left( A\varrho _{aa}^{(0)}+\kappa N\right) [2\hat{a}_{1}^{\dag
}\hat{\varrho}\hat{a}_{1}-\hat{a}_{1}\hat{a}_{1}^{\dag }\hat{\varrho}-\hat{%
\varrho}\hat{a}_{1}\hat{a}_{1}^{\dag }]+\frac{1}{2}\left( A\varrho
_{cc}^{(0)}+\kappa (N+1)\right) [2\hat{a}_{2}\hat{\varrho}\hat{a}_{2}^{\dag
}-\hat{a}_{2}^{\dag }\hat{a}_{2}\hat{\varrho}-\hat{\varrho}\hat{a}_{2}^{\dag
}\hat{a}_{2}]  \notag \\
&&+\frac{A\varrho _{ac}^{(0)}}{2}[\hat{\varrho}\hat{a}_{1}^{\dag }\hat{a}%
_{2}^{\dag }-2\hat{a}_{1}^{\dag }\hat{\varrho}\hat{a}_{2}^{\dag }+\hat{a}%
_{1}^{\dag }\hat{a}_{2}^{\dag }\hat{\varrho}-2\hat{a}_{2}\hat{\varrho}\hat{a}%
_{1}+\hat{a}_{1}\hat{a}_{2}\hat{\varrho}+\hat{\varrho}\hat{a}_{1}\hat{a}%
_{2}]-  \notag \\
&&\kappa M[\hat{a}_{1}^{\dag }\hat{\varrho}\hat{a}_{2}^{\dag }+\hat{a}%
_{2}^{\dag }\hat{\varrho}\hat{a}_{1}^{\dag }+\hat{a}_{1}\hat{\varrho}\hat{a}%
_{2}+\hat{a}_{2}\hat{\varrho}\hat{a}_{1}-\hat{a}_{1}^{\dag }\hat{a}%
_{2}^{\dag }\hat{\varrho}-\hat{\varrho}\hat{a}_{1}^{\dag }\hat{a}_{2}^{\dag
}-\hat{a}_{1}\hat{a}_{2}\hat{\varrho}-\hat{\varrho}\hat{a}_{1}\hat{a}_{2}],
\label{RO-1}
\end{eqnarray}%
where $\varrho _{ac}^{(0)}$ and $M$ are set to be real for convenience,
and $A=2r_{0}g^{2}/\gamma ^{2}$ represents the linear gain coefficient measuring the
rate at which the three-level atoms are injected into the cavity \cite{S-Z}.
In Eq. (\ref{RO-1}), the term proportional to $\varrho _{aa}^{(0)}$($\varrho
_{cc}^{(0)}$) corresponds to the gain(loss) of the mode $\mathcal{A}$($%
\mathcal{B}$), while that proportional to $\varrho _{ac}^{(0)}$ is
responsible for the intermodal coupling \cite{S-Z}. Employing Eq. (\ref{RO-1}%
) and the formula $\frac{d}{dt}\langle \mathcal{O}\rangle =\mathrm{Tr}\left(
\frac{d\hat{\varrho}}{dt}\mathcal{O}\right) $, the dynamics for the moments
of the laser modes variables turns out to be
\begin{eqnarray}
\frac{d}{dt}\langle \hat{a}_{j}(t)\rangle &=&\frac{-\Xi _{j}}{2}\langle \hat{%
a}_{j}(t)\rangle +\frac{(-1)^{j}A\varrho _{ac}^{(0)}}{2}\langle \hat{a}%
_{3-j}^{\dag }(t)\rangle \text{ \ for }j=1,2,  \label{1e-} \\
\frac{d}{dt}\langle \hat{a}_{j}^{2}(t)\rangle &=&-\Xi _{j}\langle \hat{a}%
_{j}^{2}(t)\rangle +(-1)^{j}A\varrho _{ac}^{(0)}\langle \hat{a}_{1}^{\dag
}(t)\hat{a}_{2}(t)\rangle ,  \label{2e-} \\
\frac{d}{dt}\langle \hat{a}_{j}^{\dag }(t)\hat{a}_{j}(t)\rangle &=&-\Xi
_{j}\langle \hat{a}_{j}^{\dag }(t)\hat{a}_{j}(t)\rangle +\frac{%
(-1)^{j}A\varrho _{ac}^{(0)}}{2}\left[ \langle \hat{a}_{1}^{\dag }(t)\hat{a}%
_{2}^{\dag }(t)\rangle +\langle \hat{a}_{1}(t)\hat{a}_{2}(t)\rangle \right]
+(2-j)A\varrho _{aa}^{(0)}+\kappa N,  \label{3e-} \\
\frac{d}{dt}\langle \hat{a}_{1}(t)\hat{a}_{2}(t)\rangle &=&-\frac{\Xi
_{1}+\Xi _{2}}{2}\langle \hat{a}_{1}(t)\hat{a}_{2}(t)\rangle +\frac{A\varrho
_{ac}^{(0)}}{2}\left[ \langle \hat{a}_{1}^{\dag }(t)\hat{a}_{1}(t)\rangle
-\langle \hat{a}_{2}^{\dag }(t)\hat{a}_{2}(t)\rangle +1\right] +\kappa M,
\label{4e-} \\
\frac{d}{dt}\langle \hat{a}_{1}(t)\hat{a}_{2}^{\dag }(t)\rangle &=&-\frac{%
\Xi _{1}+\Xi _{2}}{2}\langle \hat{a}_{1}(t)\hat{a}_{2}^{\dag }(t)\rangle +%
\frac{A\varrho _{ac}^{(0)}}{2}\left[ \langle \hat{a}_{1}^{2}(t)\rangle
-\langle \hat{a}_{2}^{\dag 2}(t)\rangle \right] ,  \label{5e-}
\end{eqnarray}%
where $\Xi _{1}=\kappa -A\varrho _{aa}^{(0)}$ and $\Xi _{2}=\kappa +A\varrho
_{cc}^{(0)}.$

Now, by introducing the population inversion $\eta$ defined as $\varrho
_{aa}^{(0)}=(1-\eta )/2$ with $-1\leqslant \eta \leqslant 1$ \cite{Fesseha},
and employing both $\varrho _{aa}^{(0)}+\varrho _{cc}^{(0)}=1$ and $%
|\varrho_{ac}^{(0)}|=\sqrt{\varrho _{aa}^{(0)}\varrho_{cc}^{(0)}}$, $\varrho _{cc}^{(0)}=$ $(1+\eta )/2$ and $\varrho
_{ac}^{(0)}=\varrho _{ca}^{(0)^{\ast }}=\sqrt{1-\eta ^{2}}/2$ can be readily obtained.

After setting the time derivative to zero in Eqs. [(\ref{1e-})--(\ref{5e-})],
we obtain the following non-vanishing steady-state correlations
\begin{eqnarray}
\langle \hat{a}_{1}^{\dag }\hat{a}_{1}\rangle &=&\frac{-\left[ A\eta \left(
1-\eta \right) -2\kappa \left( N-M\sqrt{1-\eta ^{2}}\right) \right] \left(
1-\eta \right) }{4\eta ^{2}\left( \kappa +A\eta \right) }  \label{s_3} \\
&&+\frac{\left( 1-\eta ^{2}\right) \left( A\eta -4\kappa N\right) +4\kappa M%
\sqrt{1-\eta ^{2}}}{2\eta ^{2}\left( 2\kappa +A\eta \right) }+\frac{\left(
N-M\sqrt{1-\eta ^{2}}\right) \left( 1+\eta \right) }{2\eta ^{2}},  \notag \\
\langle \hat{a}_{2}^{\dag }\hat{a}_{2}\rangle &=&\frac{-\left[ A\eta \left(
1-\eta \right) -2\kappa \left( N-M\sqrt{1-\eta ^{2}}\right) \right] \left(
1+\eta \right) }{4\eta ^{2}\left( \kappa +A\eta \right) }  \label{s_4} \\
&&+\frac{\left( 1-\eta ^{2}\right) \left( A\eta -4\kappa N\right) +4\kappa M%
\sqrt{1-\eta ^{2}}}{2\eta ^{2}\left( 2\kappa +A\eta \right) }+\frac{\left(
N-M\sqrt{1-\eta ^{2}}\right) \left( 1-\eta \right) }{2\eta ^{2}},  \notag \\
\langle \hat{a}_{1}\hat{a}_{2}\rangle &=&\frac{-\left[ A\eta \left( 1-\eta
\right) -2\kappa \left( N-M\sqrt{1-\eta ^{2}}\right) \right] \sqrt{1-\eta
^{2}}}{4\eta ^{2}\left( \kappa +A\eta \right) }  \label{s_5} \\
&&+\ \frac{\sqrt{1-\eta ^{2}}\left( A\eta -4\kappa N\right) +4\kappa M}{%
2\eta ^{2}\left( 2\kappa +A\eta \right) }+\frac{\left( N-M\sqrt{1-\eta ^{2}}%
\right) \sqrt{1-\eta ^{2}}}{2\eta ^{2}},
\end{eqnarray}
which are physically meaningful for only $\eta \geqslant 0$, i.e., $0\leqslant \eta \leqslant 1$.

Because the two-laser modes $\mathcal{A}$ and $\mathcal{B}$ are demonstrated to
evolve in a TMGS \cite{NDCEL}, $\hat{\varrho}_{\mathcal{AB}}$
can be fully described by the covariance matrix $\left[ \mathcal{\sigma}_{%
\mathcal{AB}}\right] _{jj^{\prime }}$ $=\langle \hat{u}_{j}\hat{u}%
_{j^{\prime }}+\hat{u}_{j^{\prime }}\hat{u}_{j}\rangle /2$, where $\hat{u}^{%
\text{\textrm{T}}}=(\hat{q}_{1},\hat{p}_{1},\hat{q}_{2},\hat{p}_{2}) $, with $%
\hat{q}_{j}=\hat{a}_{j}^{\dag }+\hat{a}_{j}$ and $\hat{p}_{j}=\mathrm{i}(%
\hat{a}_{j}^{\dag }-\hat{a}_{j})$ \cite{Shapiro}. Therefore, using Eqs. [(%
\ref{s_3})-(\ref{s_5})], $\mathcal{\sigma }_{\mathcal{AB}}$ can be expressed
as
\begin{equation}
\mathcal{\sigma }_{\mathcal{AB}}=\left( \QATOP{\alpha }{\delta ^{\mathrm{T}}}%
\QATOP{\delta }{\beta }\right) ,  \label{cm}
\end{equation}%
where $\alpha =\mathcal{\sigma }_{11}\mbox{$1 \hspace{-1.0mm}
{\bf l}$}_{2}$, $\beta =\mathcal{\sigma }_{22}%
\mbox{$1 \hspace{-1.0mm}
{\bf l}$}_{2}$ and $\delta =\mathcal{\sigma }_{12}\mathrm{diag}(1,-1)$ with $%
\mathcal{\sigma }_{jj}=2\langle \hat{a}_{j}^{\dag }\hat{a}_{j}\rangle+1 $
and $\mathcal{\sigma }_{12}=2\langle \hat{a}_{1}\hat{a}_{2}\rangle$.

\section{Gaussian quantum steering}

\label{SecIII}

First, we briefly recall what is intended by steering according \cite%
{Wiseman,Kogias}. A bipartite state $\hat{\varrho}_{\text{\textrm{AB}}}$,
shared between two observers Alice and Bob, is said to be steerable from $%
\text{A}\rightarrow \text{B}$ (i.e., Alice is able to steer Bob's states by
performing a set of measurements $\mathcal{M}_{\text{\textrm{A}}}$ on her
side), if it is impossible for every pair of local observables $\mathcal{R%
}_{\text{\textrm{A}}}\in \mathcal{M}_{\text{\textrm{A}}}$ on $\text{A}$ and $%
\mathcal{R}_{\text{\textrm{B}}}$ (arbitrary) on $\text{B}$, with respective
outcomes $r_{\text{\textrm{A}}}$ and $r_{\text{\textrm{B}}}$, to express the
joint probability as $P(r_{\text{\textrm{A}}},r_{\text{\textrm{B}}}|\mathcal{%
R}_{\text{\textrm{A}}},\mathcal{R}_{\text{\textrm{B}}},\hat{\varrho}_{\text{%
\textrm{AB}}})=\sum_{\lambda }\mathcal{P}_{\lambda }\mathcal{P}(r_{\text{%
\textrm{A}}}|\mathcal{R}_{\text{\textrm{A}}},\lambda )P(r_{\text{\textrm{B}}%
}|\mathcal{R}_{\text{\textrm{B}}},\hat{\varrho}_{\lambda })$ \cite{Wiseman},
i.e., at least one measurement pair $\mathcal{R}_{\text{\textrm{A}}}$ and $%
\mathcal{R}_{\text{\textrm{B}}}$ should violate this expression if $%
\mathcal{P}_{\lambda }$ is fixed across all measurements, where $\mathcal{P}%
_{\lambda }$ and $\mathcal{P}(r_{\text{\textrm{A}}}|\mathcal{R}_{\text{%
\textrm{A}}},\lambda )$ are\ probability distributions and $P(r_{\text{%
\textrm{B}}}|\mathcal{R}_{\text{\textrm{B}}},\hat{\varrho}_{\lambda })$
denotes the conditional probability distribution corresponding to the extra
condition of being evaluated on the state $\hat{\varrho}_{\lambda }$ \cite%
{Wiseman,Kogias}.

Within the Gaussian scenario, a necessary and sufficient condition for which
a TMGS $\hat{\varrho}_{\mathcal{XY}}$, described by its covariance matrix $%
\mathcal{\sigma }_{\mathcal{XY}}$, is steerable in the direction $\mathcal{%
X\rightarrow Y}$ under Gaussian measurements is given by $\sigma _{\mathcal{%
XY}}+i(0_{\mathcal{X}}\oplus \sigma _{\mathcal{Y}})<0$, where $0_{\mathcal{X}%
}$$=\left( \QATOP{0}{0}\QATOP{0}{0}\right) $ and $\sigma _{\mathcal{Y}}$ $%
=\left( \QATOP{0}{-1}\QATOP{1}{0}\right) $ \cite{Wiseman,Kogias}. From a
quantitative perspective, the amount by which the TMGS $\hat{\varrho}_{%
\mathcal{AB}}$ is steerable under Gaussian measurements in the direction $%
\mathcal{A\rightarrow B}$ is defined by \cite{Kogias}
\begin{equation}
\mathcal{G}^{\mathcal{A}\rightarrow \mathcal{B}}(\sigma _{\mathcal{AB}%
}):=\max \left[ 0,-\ln \left( \sqrt{\det \mathcal{S}^{\mathcal{B}}}\right) %
\right] ,  \label{S2-}
\end{equation}%
where $\mathcal{S}^{\mathcal{B}}=\beta -\delta ^{\mathrm{T}}\alpha
^{-1}\delta $ represents the Schur complement of the mode $\mathcal{A}$ in the
covariance matrix $\mathcal{\sigma }_{\mathcal{AB}}$ defined by Eq. (\ref{cm}%
).

When $\mathcal{B}$ is a single-mode party, Eq. (\ref{S2-}) acquires the
simple form \cite{Kogias}%
\begin{equation}
\mathcal{G}^{\mathcal{A}\rightarrow \mathcal{B}}=\max \left[ 0,\frac{1}{2}%
\ln \frac{\det \alpha }{\det \mathcal{\sigma }_{\mathcal{AB}}}\right],
\label{GS}
\end{equation}%
where the Gaussian $\mathcal{B}\rightarrow \mathcal{A}$ steerability can be
obtained by changing the roles of $\mathcal{A}$ and $\mathcal{B}$ in Eq. (%
\ref{GS}), i.e., $\mathcal{G}^{\mathcal{B\rightarrow A}}=\max \left[ 0,\frac{%
1}{2}\ln \frac{\det \beta }{\det \mathcal{\sigma }_{\mathcal{AB}}}\right]$. Experimentally, the three independent elements, $\mathcal{\sigma }_{11}$, $%
\mathcal{\sigma }_{22}$, and $\mathcal{\sigma }_{12}$, of the covariance
matrix (\ref{cm}), can be measured via the standard homodyne detection
method \cite{Laurat}, which allows experimental values of the
Gaussian steerabilities $\mathcal{G}^{\mathcal{A}\rightarrow\mathcal{B}}$
and $\mathcal{G}^{\mathcal{B}\rightarrow\mathcal{A}}$ to be obtained.
\begin{figure}[h]
\centerline{\includegraphics[width=0.5\columnwidth,height=5cm]{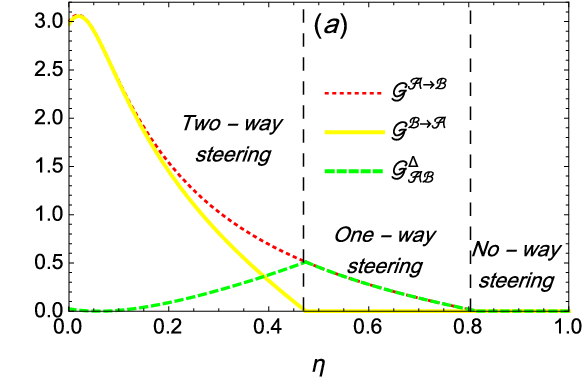}%
\includegraphics[width=0.5\columnwidth,height=5cm]{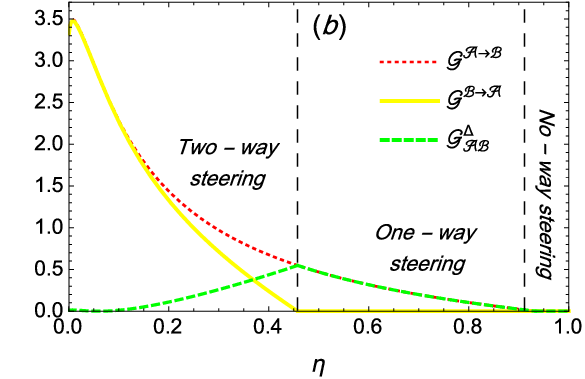}}
\caption{The Gaussian steering $\mathcal{G}^{\mathcal{A}\rightarrow \mathcal{%
B}}$, $\mathcal{G}^{\mathcal{B}\rightarrow \mathcal{A}}$ and steering
asymmetry $\mathcal{G}_{\mathcal{AB}}^{\Delta }$ versus the population
inversion $\protect\eta$ for two different values of the linear gain
coefficient $A$: (a) $A=200~\text{kHz}$ and (b) $A=1000~\text{kHz}$
corresponding to atomic injection rates $r_{0}=22~\text{kHz}$ and $r_{0}=108~%
\text{kHz}$, respectively. In both panels, we adopted $r=1.75$ for the squeezing
parameter.}
\label{f2}
\end{figure}
\newline

In the hierarchy of quantum nonseparability, EPR-steering is distinguished
from both entanglement and Bell-nonlocality by its asymmetry, i.e., the
state $\hat{\varrho}_{\mathcal{AB}}$ may be steerable in one direction, but
not vice versa \cite{Kogias}. Therefore, three scenarios are feasible: (%
\textit{i}) two-way steering, where the state $\hat{\varrho}_{\mathcal{AB}}$
is steerable in both directions, i.e., $\mathcal{G}^{\mathcal{A}\rightarrow
\mathcal{B}}>0$ and $\mathcal{G}^{\mathcal{B}\rightarrow \mathcal{A}}>0$, (%
\textit{ii}) no-way steering, where $\mathcal{G}^{\mathcal{A}\rightarrow
\mathcal{B}}=\mathcal{G}^{\mathcal{B}\rightarrow \mathcal{A}}=0$, and (%
\textit{iii}) one-way steering, in which the steerability is authorized only
in one direction, i.e., ($\mathcal{G}^{\mathcal{A}\rightarrow \mathcal{B}}>0$
with $\mathcal{G}^{\mathcal{B}\rightarrow \mathcal{A}}=0$) or ($\mathcal{G}^{%
\mathcal{A}\rightarrow \mathcal{B}}=0$ with $\mathcal{G}^{\mathcal{B}%
\rightarrow \mathcal{A}}>0$). Note that pure entangled states cannot
exhibit one-way steering, since they can always be transformed into a
symmetric form via a local basis change using the Schmidt decomposition \cite%
{Kogias}. Finally, to check how asymmetric can the steerability be in the
state $\hat{\varrho}_{\mathcal{AB}} $, we study the steering asymmetry
defined by $\mathcal{G}_{\mathcal{AB}}^{\Delta }\doteq\left\vert \mathcal{G}%
^{\mathcal{A}\rightarrow \mathcal{B}}-\mathcal{G}^{\mathcal{B}\rightarrow
\mathcal{A}}\right\vert $ \cite{Kogias}.

To test our protocol against experimental capabilities, we use parameters obtained from \cite{Han,Meschede}: the cavity decay rate for both laser modes $%
\mathcal{A}$ and $\mathcal{B}$ $\kappa=3.85~\text{kHz}$, atomic decay
rate $\gamma=20~\text{kHz}$, coupling constant $g=43~\text{kHz}$, and $%
r_{0}=22~\text{kHz}$ for the rate at which the atoms are initially injected
into the cavity. With these parameters, we obtain $A=2r_{0}g^{2}/\gamma
^{2}\approx200~\text{kHz}$ for the linear gain coefficient. In Fig. \ref{f2}%
, we plot the steerabilities $\mathcal{G}^{\mathcal{A}\rightarrow \mathcal{B}%
}$ and $\mathcal{G}^{\mathcal{B}\rightarrow \mathcal{A}}$ and steering asymmetry $\mathcal{G}_{\mathcal{AB}}^{\Delta }$ against the
population inversion $\eta $ for two different values of the linear gain
coefficient $A$. It is not difficult to realize that inferring on mode $%
\mathcal{B}$ based on the measurements performed on mode $\mathcal{A}$ completely differs from the reverse process, which is an interesting
example of the role played by the observer in quantum mechanics.

Interestingly, Fig. \ref{f2} illustrates that when $\eta \rightarrow 1$, i.e., $%
\varrho _{cc}^{(0)}\approx 1$ and $\varrho _{aa}^{(0)}\approx \varrho
_{ac}^{(0)}\approx 0$, $\mathcal{G}^{\mathcal{A}\rightarrow \mathcal{B}}=%
\mathcal{G}^{\mathcal{B}\rightarrow \mathcal{A}}=0$, regardless of the value
of $A$. This result asserts that if the atoms are initially prepared in the
lower energy level, there is no possibility that they emit a radiation
capable of forging quantum steering between modes $\mathcal{A}$ and $%
\mathcal{B}$. In contrast, when $\eta \rightarrow 0$, i.e., $\varrho
_{aa}^{(0)}\approx \varrho _{cc}^{(0)}\approx \varrho _{ac}^{(0)}\approx 1/2$, which corresponds to a maximum initial atomic coherence, the state $\hat{%
\varrho}_{\mathcal{AB}}$ is maximally steerable in both directions $\mathcal{%
A}\rightarrow \mathcal{B}$ and $\mathcal{B}\rightarrow \mathcal{A}$. In addition, we remark that by increasing the population inversion $\eta $, two-way
steering can be quickly transformed into one-way steering using larger values of $%
A$. Remarkably, Fig. \ref{f2} illustrates the existence of situations in which the
state $\hat{\varrho}_{\mathcal{AB}}$ can be steered only from $\mathcal{A}%
\rightarrow \mathcal{B}$. This could be interpreted as follows: among their
shared state $\hat{\varrho}_{\mathcal{AB}}$, Alice and Bob can perform the
same Gaussian measurements, but obtain different outputs. In other words,
Alice who possesses mode $\mathcal{A}$ can convince Bob who has mode $%
\mathcal{B}$ that their shared state is entangled, whereas, the reverse
process is impossible.
\begin{figure}[h]
\centerline{\includegraphics[width=0.5\columnwidth,height=5cm]{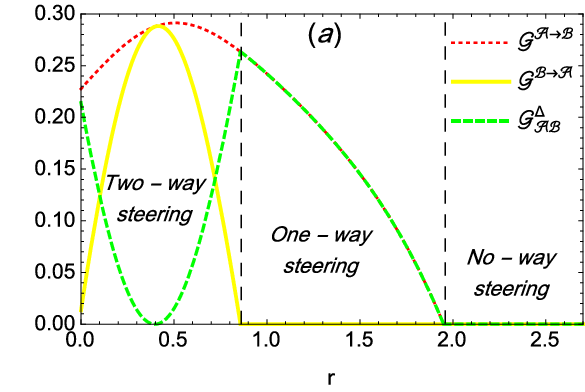}%
\includegraphics[width=0.5\columnwidth,height=5cm]{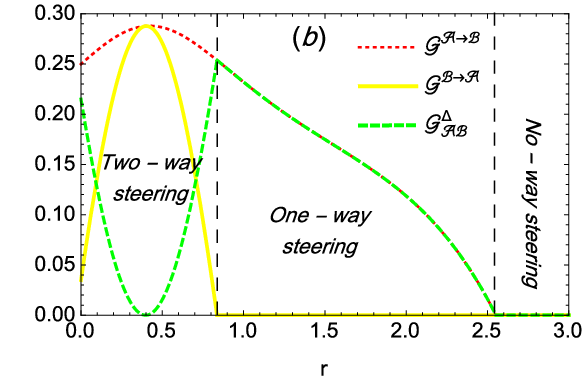}}
\caption{The Gaussian steering $\mathcal{G}^{\mathcal{A}\rightarrow \mathcal{%
B}}$, $\mathcal{G}^{\mathcal{B}\rightarrow \mathcal{A}}$ and steering
asymmetry $\mathcal{G}_{\mathcal{AB}}^{\Delta }$ against the squeezing
parameter $r$ using $A=200~\text{kHz}$ in (a) and $A=1000~\text{kHz}$ in
(b). The population inversion is fixed as $\protect\eta =0.75$ in both
panels. }
\label{f3}
\end{figure}
\newline

In Fig. \ref{f3}, we plot the steerabilities $\mathcal{G}^{\mathcal{A}%
\rightarrow \mathcal{B}}$, $\mathcal{G}^{\mathcal{B}\rightarrow \mathcal{A}}$
and steering asymmetry $\mathcal{G}_{\mathcal{AB}}^{\Delta }$ against the
squeezing parameter $r$. The linear gain coefficient $A$ is the same as in
Fig. \ref{f2}. Similar to the results presented in Fig. \ref{f2}, Fig. \ref%
{f3} shows that, under realistic experimental conditions, one-way steering
still emerges from $\mathcal{A}\rightarrow\mathcal{B}$, which stems from the
fact that the $\mathcal{A}\rightarrow \mathcal{B}$ steerability remains
regularly more than that from $\mathcal{B}\rightarrow \mathcal{A}$. In
particular, Figs. \ref{f2}(b) and \ref{f3}(b) clearly illustrate the advantage of
the mediation of the linear gain coefficient $A $ in observing one-way
steering via a wide range of the parameters $\eta $ and $r$, which
allows us to achieve more one-way quantum information tasks. Furthermore, as can be observed from Fig. \ref{f3}(b), both steerabilities $\mathcal{G}^{\mathcal{A}%
\rightarrow \mathcal{B}}$ and $\mathcal{G}^{\mathcal{B}\rightarrow \mathcal{A%
}}$ are non-zero, even for $r=0$, which implies that owing to atomic coherence (i.e.,
$\eta \neq 1$), quantum steering between the modes $\mathcal{A}$ and $%
\mathcal{B}$ can be created without squeezed vacuum light.

Quite remarkably, the steerabilities $\mathcal{G}^{\mathcal{A}\rightarrow
\mathcal{B}}$ and $\mathcal{G}^{\mathcal{B}\rightarrow \mathcal{A}}$ undergo
a resonance-like behavior under squeezing effect. This can be explained as
follows: because the reduced state of two-mode squeezed noise is a thermal
state with a mean photon number proportional to the squeezing parameter $r$
\cite{Shapiro}, with increasing $r$, the number of photons
entering the cavity also increases, which induces additional coherence and
further enhances the steering in both directions. In contrast, after
reaching their maximum, $\mathcal{G}^{\mathcal{A}\rightarrow \mathcal{B}}$
and $\mathcal{G}^{\mathcal{B}\rightarrow \mathcal{A}}$ start to decrease
under the decoherence effect triggered by high squeezing $r$. More importantly, Fig. %
\ref{f3} indicates that squeezed noise can play a constructive role in
realizing one-way steering, which originates from the fact that under squeezing effect, the steering $%
\mathcal{G}^{\mathcal{B}\rightarrow \mathcal{A}}$ decays more rapidly to zero than $\mathcal{G}^{\mathcal{A}\rightarrow \mathcal{B}}$. Here, it is worth noting that two-mode squeezed states with strong
squeezing up to 10~\textrm{dB}(i.e., $r>1$) are well within experimental
reach, as demonstrated in \cite{Vahlbruch}.

In miscellaneous TMGSs $\hat{\varrho}_{\mathcal{XY}}$ \cite%
{LPL,SOPTO,magno,LBS,ElQars,PRA,kongHu}, it has been demonstrated that one-way
steering behavior can be detected from both $\mathcal{X\rightarrow Y}$ and $\mathcal{Y\rightarrow X}$. However, under various realistic experimental circumstances, such behavior solely emerges from $\mathcal{A}%
\rightarrow \mathcal{B}$ in the considered state $\hat{\varrho}_{\mathcal{AB}}$%
, as vividly illustrated in Figs. \ref{f2} and \ref{f3}. Next, we
will demonstrate that the state $\hat{\varrho}_{\mathcal{AB}}$ can, in general,
exhibit one-way steering solely from $\mathcal{A}\rightarrow \mathcal{B}$, regardless of its physical and environmental parameters. \newline
\begin{figure}[h]
\centerline{\includegraphics[width=0.5\columnwidth,height=6cm]{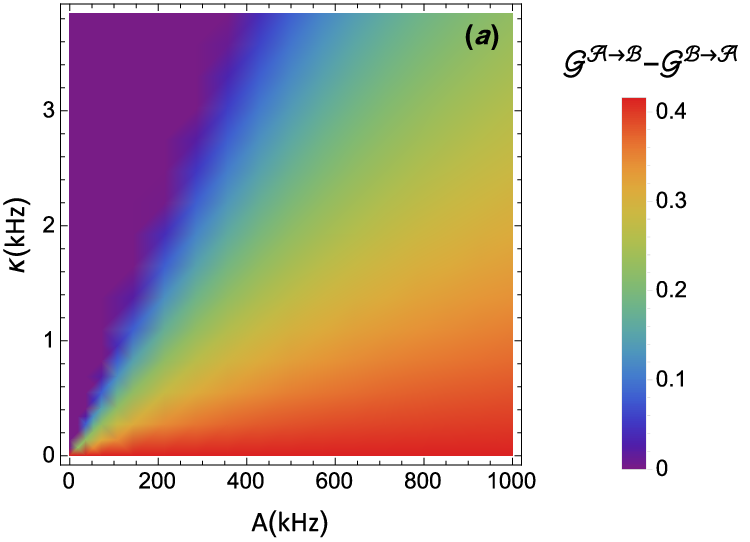}%
\includegraphics[width=0.5\columnwidth,height=6cm]{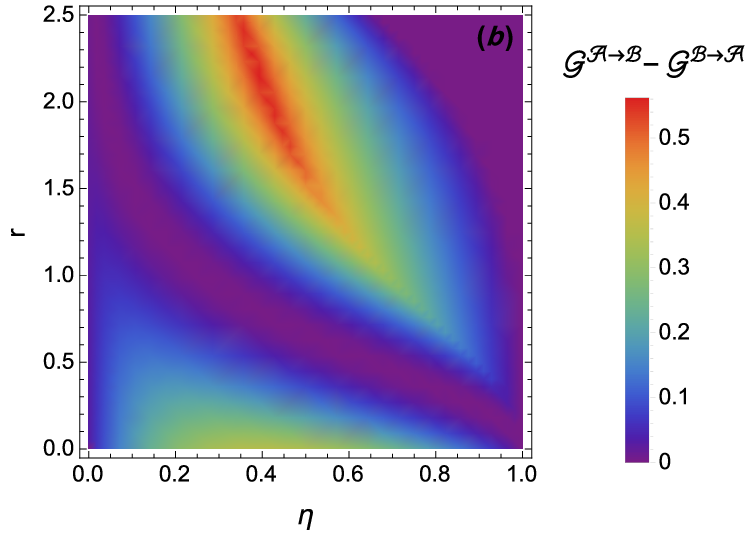}}
\caption{(a) Density plot of $\mathcal{G}^{\mathcal{A}\rightarrow \mathcal{B}%
}-\mathcal{G}^{\mathcal{B}\rightarrow \mathcal{A}}$ versus $\protect\kappa $
and $A$ using $\protect\eta =0.5$ and $r=2.75$. (b) Density plot of $%
\mathcal{G}^{\mathcal{A}\rightarrow \mathcal{B}}-\mathcal{G}^{\mathcal{B}%
\rightarrow \mathcal{A}}$ versus $r$ and $\protect\eta $ using $\protect%
\kappa =3.85~\text{kHz}$ and $A=500~\text{kHz}$ (corresponding to an
injection rate $r_{0}=54~\text{kHz}$).}
\label{f4}
\end{figure}
\newline

Using density values of the parameters $\eta ,$ $\kappa ,$ $A$, and $r$,
Fig. \ref{f4} demonstrates that the difference $\mathcal{G}^{\mathcal{A}\rightarrow
\mathcal{B}}-\mathcal{G}^{\mathcal{B}\rightarrow \mathcal{A}}$ remains
positive. Therefore, achieving $\mathcal{G}^{\mathcal{A}\rightarrow \mathcal{%
B}}=0$ and $\mathcal{G}^{\mathcal{B}\rightarrow \mathcal{A}}>0$
simultaneously is forbidden under such circumstances. Owing to this fact,
observing one-way steering in the direction $\mathcal{B}\rightarrow \mathcal{%
A}$ is impossible, which we now demonstrate to be free from the choice of $%
\eta ,$ $\kappa ,$ $A$, and $r$. In fact, because one-way steering from $%
\mathcal{A}\rightarrow \mathcal{B}$ requires both $\mathcal{G}^{\mathcal{A}%
\rightarrow \mathcal{B}}>0$ and $\mathcal{G}^{\mathcal{B}\rightarrow
\mathcal{A}}=0$, which originates from $\mathcal{G}^{\mathcal{A}\rightarrow
\mathcal{B}}\geqslant \mathcal{G}^{\mathcal{B}\rightarrow \mathcal{A}}$, it
follows that the general holding of the constraint $\mathcal{G}^{\mathcal{A}%
\rightarrow \mathcal{B}}\geqslant \mathcal{G}^{\mathcal{B}\rightarrow
\mathcal{A}}$ imposes the state $\hat{\varrho}_{\mathcal{AB}}$ to display
one-way steering behavior only from $\mathcal{A}\rightarrow \mathcal{B}$. However, by employing Eqs. (\ref{cm}) and (\ref{GS}), it can be deduced that $%
\mathcal{G}^{\mathcal{A}\rightarrow \mathcal{B}}\geqslant \mathcal{G}^{%
\mathcal{B}\rightarrow \mathcal{A}}$ is equivalent to $\langle \hat{a}%
_{1}^{\dag }\hat{a}_{1}\rangle \geqslant \langle \hat{a}_{2}^{\dag }\hat{a}%
_{2}\rangle $, which is closely related to the intensity difference $\mathcal{%
I}=\langle \hat{a}_{1}^{\dag }\hat{a}_{1}\rangle -\langle \hat{a}_{2}^{\dag }%
\hat{a}_{2}\rangle $, where $\langle \hat{a}_{1}^{\dag }\hat{a}_{1}\rangle$($%
\langle \hat{a}_{2}^{\dag }\hat{a}_{2}\rangle $) represents the mean photon number
in the mode $\mathcal{A}$($\mathcal{B}$) \cite{Milburn}. Next, using Eqs. (%
\ref{s_3}) and (\ref{s_4}), we obtain
\begin{equation}
\mathcal{I}=\frac{A\left( 1-\eta +2\sinh ^{2}\left( r\right) -2\sinh \left(
r\right) \cosh \left( r\right) \sqrt{1-\eta ^{2}}\right) }{2\left( \kappa
+A\eta \right) }.  \label{SN}
\end{equation}

First, we remark that the intensity difference $\mathcal{I}$ vanishes
for $A=0$, as there are no atoms inside the cavity and therefore no laser
emission. However, owing to the squeezed noise fluctuations entering the
cavity ($r>0$), $\mathcal{I}$ could be non-zero, even without atomic
coherence (i.e., when $\eta =1$). Importantly, using the fact that $\left(
\sqrt{1+\eta }\sinh (r)-\sqrt{1-\eta }\cosh (r)\right) ^{2}\geqslant 0$ and $%
\cosh ^{2}(r)-\sinh ^{2}(r)=1$, it can be verified that $\mathcal{I}$ is always
positive, thus implying that the constraint $\mathcal{G}^{\mathcal{A}\rightarrow
\mathcal{B}}\geqslant \mathcal{G}^{\mathcal{B}\rightarrow \mathcal{A}}$ is
generally fulfilled by the state $\hat{\varrho}_{\mathcal{AB}}$. Consequently, this guarantees one-way steering to be, \textit{in general},
unidirectional from $\mathcal{A}\rightarrow \mathcal{B}$ in the state $\hat{%
\varrho}_{\mathcal{AB}}$, regardless of the parameters $\eta $, $\kappa $, $A
$, and $r$. The demonstrated unidirectional one-way steering may offer
practical implications for the distribution of the trust in asymmetric
quantum information processing, e.g., quantum communication networks \cite%
{OWS,QKD,kongHu}.

Finally, Figs. \ref{f2} and \ref{f3} show that the steering asymmetry $%
\mathcal{G}_{\mathcal{AB}}^{\Delta }$ cannot exceed $\ln 2=0.69$, is
maximal when the state $\hat{\varrho}_{\mathcal{AB}}$ is nonsteerable in one
direction, and decreases with increasing steerability in both directions $%
\mathcal{A}\rightarrow \mathcal{B}$ and $\mathcal{B}\rightarrow \mathcal{A}$%
. In other words, the attainment of maximal steering asymmetry $\mathcal{G}_{%
\mathcal{AB}}^{\Delta }$ witnesses the transition between one-way and
two-way steering in the TMGS $\hat{\varrho}_{\mathcal{AB}} $. In addition,
under various realistic experimental conditions, Fig. \ref{f4} shows that
the difference $\mathcal{G}^{\mathcal{A}\rightarrow \mathcal{B}}-\mathcal{G}%
^{\mathcal{B}\rightarrow \mathcal{A}}$ is bounded from below and above by $0$
and $0.61$, respectively, thus implying that the steering asymmetry $\mathcal{G}%
_{\Delta }^{\mathcal{AB}}\doteq |\mathcal{G}^{\mathcal{A}\rightarrow
\mathcal{B}}-\mathcal{G}^{\mathcal{B}\rightarrow \mathcal{A}}|$ cannot
exceed $\ln 2=0.69$, which implies that the TMGS $\hat{\varrho}_{\mathcal{AB}}$
never diverges to an extremal asymmetry state.

With the availability of strong squeezing up to $10~\text{dB}$ \cite%
{Vahlbruch}, and the combination of cavity-quantum electrodynamics and
trapped atoms techniques \cite{Rempe,QED}, together with the standard
homodyne detection method \cite{Laurat}, our efficient
unidirectional Gaussian one-way steering may be realized experimentally.

\section{Conclusion}

\label{SecIV}

In this work, we studied asymmetric steering in a TMGS $\hat{\varrho}_{%
\mathcal{AB}}$ coupled to a two-mode squeezed vacuum reservoir. The modes $%
\mathcal{A}$ and $\mathcal{B}$ were generated via the upper
and lower transitions of a nondegenerate three-level laser, respectively. Within the
linear-adiabatic approximation, the master equation of the state $\hat{%
\varrho}_{\mathcal{AB}}$ was obtained and then used to derive its
corresponding covariance matrix $\sigma_{\mathcal{AB}}$. Under realistic
experimental conditions, we demonstrated that asymmetric steering can be generated
in the state $\hat{\varrho}_{\mathcal{AB}}$ with maximum steering asymmetry
less than $\ln 2$, which implies that the state $\hat{\varrho}_{\mathcal{AB}}$
never evolves to an extremal asymmetry state. Moreover, we determined that the $%
\mathcal{B}\rightarrow\mathcal{A}$ steerability is always less than
that from $\mathcal{A}\rightarrow\mathcal{B}$, and that it decays rapidly to zero
under squeezing effect. Despite being a source of decoherence effect, we demonstrated that squeezed noise can play a positive role in realizing one-way
steering in the studied state.

Although one-way steering is observed from both $\mathcal{X}\rightarrow\mathcal{Y}$ and $\mathcal{Y}\rightarrow\mathcal{X}$ in various TMGSs $\hat{%
\varrho}_{\mathcal{XY}}$ \cite{LPL,SOPTO,magno,LBS,ElQars,PRA,kongHu}, we demonstrated that the state $\hat{\varrho}_{\mathcal{AB}}$ can, \textit{in general}%
, exhibit one-way steering solely from $\mathcal{A\rightarrow B}$. Such
unidirectionality restriction is shown to emerge as a consequence of the
fact that the intensity difference of the modes $\mathcal{A}$ and $\mathcal{B%
}$ is verified to remain always positive, regardless of the physical and
environmental parameters of $\hat{\varrho}_{\mathcal{AB}}$.

With these results, we believe that nondegenerate three-level lasers may
offer interesting perspectives in the realization of genuine one-way
steering, which may provide a useful resource for the distribution of the
trust, e.g., in quantum communication networks. Finally, it would be
interesting to find an application of the unidirectional
one-way steering phenomenon in future one-way quantum information tasks.

\end{document}